%% file: main.tex
\documentclass[conference]{IEEEtran}
\usepackage{cite}
\usepackage{amsmath,amssymb,amsfonts}
\usepackage{algorithmic}
\usepackage{graphicx}
\usepackage{textcomp}
\usepackage{xcolor}
\usepackage{url}
\usepackage{multirow}

\usepackage{subfigure}
\usepackage{enumitem}
\usepackage{fancyvrb}
\usepackage{pifont}
\newcommand{\cmark}{\ding{51}}%
\newcommand{\xmark}{\ding{55}}%

\newcommand{\ourmethod}{PoFT}

\def\BibTeX{{\rm B\kern-.05em{\sc i\kern-.025em b}\kern-.08em
    T\kern-.1667em\lower.7ex\hbox{E}\kern-.125emX}}
\begin{document}

\title{
\textit{Proof of Federated Training}: Accountable Cross-Network Model Training and Inference
}

\author{\IEEEauthorblockN{Sarthak Chakraborty\IEEEauthorrefmark{2}\IEEEauthorrefmark{1},
Sandip Chakraborty\IEEEauthorrefmark{3},
}
\IEEEauthorblockA{\IEEEauthorrefmark{2}\textit{Adobe Research, Bangalore}, \IEEEauthorrefmark{3}\textit{Indian Institute of Technology, Kharagpur}  \\
sarthak.chakraborty@gmail.com, sandipc@cse.iitkgp.ac.in}
\thanks{\IEEEauthorrefmark{1}Work done while at Indian Institute of Technology, Kharagpur}
}


\IEEEoverridecommandlockouts
\IEEEpubid{\makebox[\columnwidth]{978-1-6654-9538-7/22/\$31.00~\copyright2022 IEEE \hfill} \hspace{\columnsep}\makebox[\columnwidth]{ }}

\maketitle


\begin{abstract}
Blockchain has widely been adopted to design accountable federated learning frameworks; however, the existing frameworks do not scale for distributed model training over multiple independent blockchain networks. For storing the pre-trained models over blockchain, current approaches primarily embed a model using its structural properties that are neither scalable for cross-chain exchange nor suitable for cross-chain verification. This paper proposes an architectural framework for cross-chain verifiable model training using federated learning, called \textit{Proof of Federated Training} (\ourmethod), the first of its kind that enables a federated training procedure span across the clients over multiple blockchain networks. Instead of structural embedding, \ourmethod{} uses model parameters to embed the model over a blockchain and then applies a verifiable model exchange between two blockchain networks for cross-network model training. We implement and test \ourmethod{} over a large-scale setup using Amazon EC2 instances and observe that cross-chain training can significantly boosts up the model efficacy. In contrast, \ourmethod{} incurs marginal overhead for inter-chain model exchanges. 
\end{abstract}


\input{tex/01.Intro}

\input{tex/02.Background_new}

\input{tex/04.System}

\input{tex/05.Implementation}

\input{tex/06.Evaluation}

\input{tex/07.Conclusion}

\bibliographystyle{IEEEtran}
\bibliography{main}


\end{document}

%% file: tex/01.Intro.tex
\section{Introduction}

Federated Learning (FL) frameworks~\cite{lo2021systematic} have widely been deployed in various large-scale networked systems like Google Keyboard, Nvidia Clara Healthcare Application, etc., employing distributed model training over locally preserved data, thus supporting data privacy~\cite{cheng2020federated}. A typical FL setup proceeds in rounds where individual clients fetches a global model to train them locally and independently, consuming their own data sources. The clients then forward these local models to a server that aggregates them using an aggregation strategy and updates the global model, and this entire procedure runs in iteration. Such a framework is useful when multiple organizations need to collectively train a Deep Neural Network (DNN) model without explicitly sharing their local data~\cite{niu2020billion,zhong2021p}. However, FL is prone to a wide range of security attacks~\cite{ma2020safeguarding,jere2020taxonomy,desai2021blockfla,fang2020local}. Consequently, different works~\cite{goel2019deepring,awan2019poster,ramanan2020baffle,peng2021vfchain,feng2021blockchain} have developed accountable FL architectures where different versions of the models, along with the model execution steps, are audited over a distributed public ledger, primarily a permissioned blockchain-based system~\cite{sharma2019blurring} to make the model training stages accountable and verifiable. In a blockchain-based decentralized FL architecture, the clients collectively execute the server as a \textit{service over the blockchain}. The local models from the client, along with the global models generated at each iteration of the FL, are recorded over the blockchain, ensuring accountability of the models by letting clients verify any of the local/global models with public test data~\cite{peng2021vfchain}.

However, the existing approaches for decentralized FL over blockchain do not scale when the organizations running the FL clients are part of multiple independent blockchain networks. With blockchain networks running in a silo, the organizations can subscribe to one or more such networks to obtain specific services. For example, several application services, like \textit{MedicalChain}\footnote{\url{https://medicalchain.com/} (Access: \today)}, \textit{Coral Health}\footnote{\url{https://mycoralhealth.com/product/} (Access: \today)}, \textit{Patientory}\footnote{\url{https://patientory.com/} (Access: \today)}, etc. use their individual blockchain networks to store patient data and apply deep learning techniques to process the data over the blockchains. Interestingly, these networks contain data from a similar domain (e.g., patients' medical imaging data), and the features learned can be exploited to develop a rich disease diagnosis model by a service like \textit{Clara Medical Imaging}\footnote{\url{https://developer.nvidia.com/clara} (Access: \today)}.

\begin{figure}[t]
        \centering 
        \includegraphics[width=\linewidth,keepaspectratio]{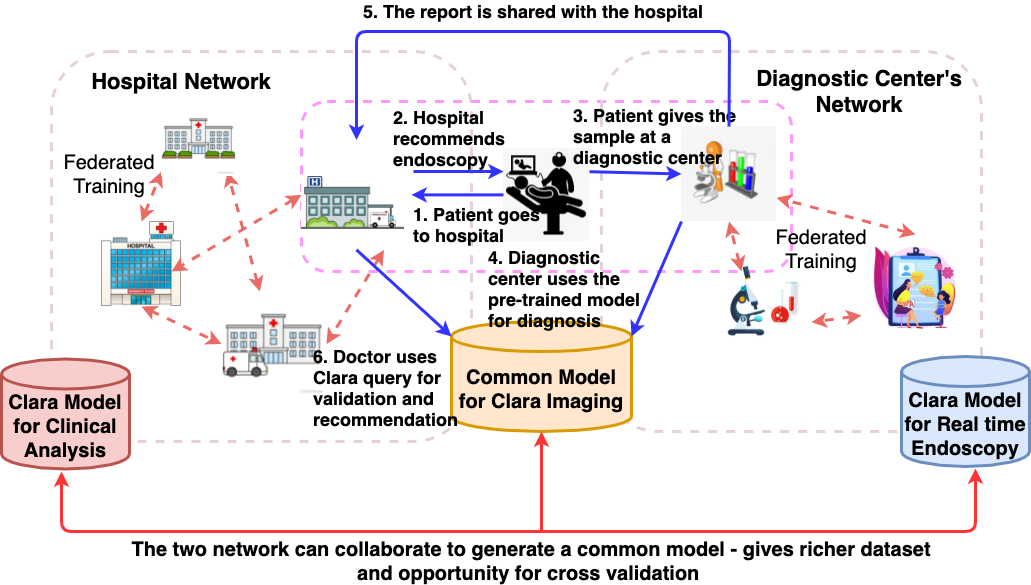}%
        \caption{An Example Use-case of \ourmethod}
        \label{fig:usecase}%
\end{figure}

A concrete use case where FL over multiple blockchains can be used in practice exists in medical domain~\cite{courtiol2019deep} (shown in Fig. \ref{fig:usecase}), where a group of hospitals use Clara Medical Imaging over a Blockchain-based distributed ledger network\footnote{https://blogs.nvidia.com/blog/2019/12/01/clara-federated-learning/} to train and use a model for a personalized recommendation to the doctors based on clinical symptoms. Let another group of diagnostic centers use Clara for real-time endoscopy, and these diagnostic centers form another blockchain network. Now, patients may visit the diagnostic center on recommendation from the hospital, whereby the endoscopy imaging data can be shared between the diagnostic center and the network of hospitals (by maintaining appropriate privacy and data accountability via FL, as used in Clara) to make the model learn better and assist the hospital doctors in clinical diagnosis. However, there is a reluctance to share the complete internal data with each other across the silos, but wish to share only parts of the entire data. The open research question is \textit{how can we help the Clara FL model to get trained over both the networks jointly?}

Not only in medical domain, similar use cases exist in agri-tech~\cite{durrant2022role} where a model is trained to predict yield production in a given season and the raw data is usually not transferred across silos. Use cases in banking sector~\cite{yang2019ffd} also has a substantial market in the domain of cross-silo federated learning, where a credit card fraud detection is one of the major conundrum. 

To share a model among multiple blockchain networks, the primary requirements to be satisfied are as follows. (1) Every blockchain network should be able to independently verify individual versions of the global model (for an FL setup) received from another blockchain network without any dependency on previous versions. (2) The cross-network transfer and in-network update of the global model versions should run asynchronously. (3) The cross-network transfer protocol should be Byzantine-safe by design to prevent clients from exhibiting Byzantine behavior. Interestingly, the first two requirements are not satisfied by the existing blockchain-based decentralized FL frameworks~\cite{goel2019deepring,awan2019poster,ramanan2020baffle,peng2021vfchain,feng2021blockchain} that use a structural representation of the model by storing the layers and activations as assets over the blockchain. When DNN structures change, multiple assets of previous versions need to be transferred across networks for cross-chain verifiability, since a blockchain asset storing structural representation is not verifiable independently, which is not feasible in practice. Although blockchain interoperability and cross-chain asset transfer protocols~\cite{jin2018towards,abebe2019enabling,syta2016keeping,belchior2020survey,liu2019hyperservice,ghosh2021leveraging} address the third requirement as mentioned above, they do not ensure distributed control over the model training and transparency in the training process that will help prevent attacks such as model poisoning. 

This paper proposes \textit{Proof of Federated Training}\footnote{Here, we do not use the notion of `proof' similar to a consensus algorithm like PoW or PoS; We provide an audtibale system and hence, the apellation.} (\ourmethod{}), the first of its kind cross-chain scalable federated training framework that can work over multiple blockchain networks. \ourmethod{} framework decouples model update from model verification; thus with asynchronous updating of models and running of blockchain services. We design a verifiable representation of DNNs as learning assets over a blockchain, which can seamlessly be transferred from one network to another without having any dependency on the model update stages or versioning of the global models. Further, \ourmethod{} utilizes the model parameters (the \textit{weight} vector) to represent the learning assets replacing the structural embedding of the model while ensuring its standalone verifiability. Finally, \ourmethod{} provides a method for cross-chain transfer and verification of the learning assets, enabling the clients of different blockchain networks to update their local models as well as the corresponding global model version based on the pre-trained global models from other networks. We implement a large-scale test network over Amazon AWS to analyze the performance of \ourmethod{} with an image classification task using $4$ different DNN models. From a thorough analysis of the models with a varying number of clients ($10$ to $40$) under each blockchain network, we observe that \ourmethod{} is scalable. It is comforting to see that even for huge DNN models like \textit{Residual Network} (\textit{ResNet-18}) having more than $10$million parameters, \ourmethod{} can complete each round of model updates within a few minutes, whereas transferring and verification of the model from one network to another take $\sim0.5$ min and $\sim2.5$min, respectively.  


%% file: tex/02.Background_new.tex
\section{Related Work} \label{sec:background}

The primary components that form the backbone of \ourmethod{} are based on Federated Learning~\cite{mcmahan2017communication, bonawitz2019towards} and blockchain~\cite{belotti2019vademecum}. FL has been widely adopted in practice for use cases where the data resides with individuals, but a machine learning model is trained in a distributed fashion. Being a distributed mode of training, the accountability and trustworthiness of individual data sources remain a question. As blockchain~\cite{belotti2019vademecum} provides a secure and trustworthy decentralized public ledger platform for data and asset sharing, a large number of existing works have adopted the blockchain technology to design frameworks for accountable FL~\cite{kim2018device,korkmaz2020chain,lo2021blockchain}. However, these works use blockchain as a separate service over FL, which causes latency issues. Also, they target a single blockchain framework, and hence no verifiable learning assets are designed that can be transferred across silos. A few more recent works, as listed in Table \ref{tab:related_works}, though addresses some of these shortcomings, including improving latency, are still not sufficient for cross-chain model transfer.

\vspace{-1mm}
\begin{table}[h]
\centering
\caption{Comparison of previous works}
\vspace{-1mm}
\begin{tabular}{|c|cccc|}
\hline
\multirow{2}{*}{} & \multicolumn{4}{c|}{Accountable FL} \\ \cline{2-5} 
 &  \multicolumn{1}{c|}{\begin{tabular}[c]{@{}c@{}}Private\\ Blockchain\end{tabular}} & \multicolumn{1}{c|}{\begin{tabular}[c]{@{}c@{}}Verifiable\\ Assets\end{tabular}} & \multicolumn{1}{c|}{\begin{tabular}[c]{@{}c@{}}Independent\\ Verification\end{tabular}} & \begin{tabular}[c]{@{}c@{}}Cross-\\ Chain FL\end{tabular} \\ \hline
 
BlockFLA~\cite{desai2021blockfla} &  \multicolumn{1}{c|}{\xmark} & \multicolumn{1}{c|}{\xmark}  & \multicolumn{1}{c|}{\xmark}  & \xmark \\ \hline

Deepring~\cite{goel2019deepring} &  \multicolumn{1}{c|}{\cmark} & \multicolumn{1}{c|}{\xmark} & \multicolumn{1}{c|}{\xmark}  & \xmark \\ \hline

Deepchain~\cite{weng2019deepchain} &  \multicolumn{1}{c|}{\cmark} & \multicolumn{1}{c|}{\xmark}  & \multicolumn{1}{c|}{\xmark}  & \xmark \\ \hline

Baffle\cite{ramanan2020baffle} &  \multicolumn{1}{c|}{\xmark} & \multicolumn{1}{c|}{\cmark} & \multicolumn{1}{c|}{\cmark}  & \xmark \\ \hline

VFchain~\cite{peng2021vfchain} &  \multicolumn{1}{c|}{\cmark} & \multicolumn{1}{c|}{\cmark} & \multicolumn{1}{c|}{\xmark}  & \xmark \\ \hline

\textbf{\ourmethod{}} &  \multicolumn{1}{c|}{\cmark} & \multicolumn{1}{c|}{\cmark}  & \multicolumn{1}{c|}{\cmark}  & \cmark \\ \hline

\end{tabular}
\label{tab:related_works}
\end{table}
\vspace{-1mm}

However, none of these works are targeted for cross-chain federated training. With individual enterprises operating on different blockchain platforms in silos, there is a need for interoperation among these respective networks. Hence, a trusted system of cross-chain interoperation involving multiple blockchains is deemed necessary.

Blockchain interoperability and cross-chain asset transfer have also been focused on in the recent literature~\cite{belchior2020survey}. These methodologies try to transfer operation flow by allowing clients of a separate network to download the asset and store them in their own local ledger. It is easier to transfer assets within two public blockchains since any client can join the network and alter the state of the blockchain. Heterogeneous blockchain interoperability has been studied in ~\cite{liu2019hyperservice, ghosh2021leveraging}. IBM has used relay service and additional smart contracts~\cite{abebe2019enabling} to verify the document transferred among permissioned blockchains. Cryptographic signature mechanisms to digitally sign the documents like Collective Signature~\cite{syta2016keeping} ensure verifiability. However, their system modeling does not optimize the design to enable cross-chain federated training for learning a model. Thus, our objective varies from the above works to optimize the interoperability architecture between two permissioned blockchains to train a learning model collectively after the transfer of assets.

%% file: tex/04.System.tex
\section{Problem Statement and Solution Overview}
\label{sec:overview}
Let us define a formal setting where there be two independent networks $\mathcal{N}_1$ and $\mathcal{N}_2$ that independently maintain their own permissioned blockchain networks, $\mathcal{B}_1$ and $\mathcal{B}_2$, respectively, to train DL models using a decentralized FL framework. Each network has a set of enterprises $\{E_{\mathcal{N}_1}^1, E_{\mathcal{N}_1}^2, \hdots\} \in \mathcal{N}_1$ and $\{E_{\mathcal{N}_2}^1, E_{\mathcal{N}_2}^2, \hdots\} \in \mathcal{N}_2$ that run the FL client service over the respective blockchain of their networks. The precise objective of \ourmethod{} is to support interoperability between $\mathcal{B}_1$ and $\mathcal{B}_2$, such that the assets containing global model versions can be shared among the clients of $\mathcal{N}_1$ and $\mathcal{N}_2$ to develop an aggregated global model while utilizing the rich volume of data available across all clients in $\mathcal{N}_1$ and $\mathcal{N}_2$.

\begin{figure}[!ht]
    \centering 
    \includegraphics[width=\linewidth]{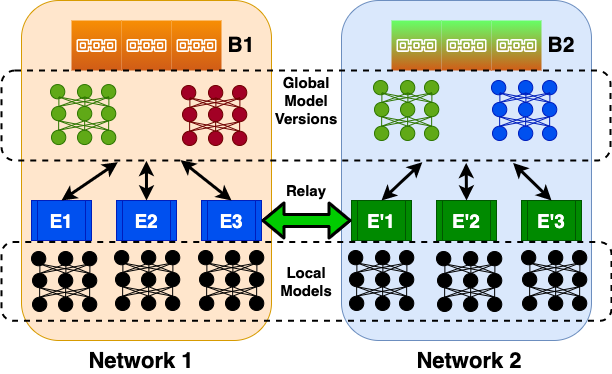}
    \caption{Federated Training of Neural Networks Across Two Blockchains}
    \label{fig:PoTF}
\end{figure}

\subsection{Why Do Existing Models Fail?}
To support cross-network model training using existing approaches, a general approach is to independently train the model partially over a network (say $\mathcal{N}_1$), and then transfer the partially trained models from $\mathcal{N}_1$ to $\mathcal{N}_2$ using existing blockchain interoperability solutions, such as~\cite{abebe2019enabling}. However, how can we represent the partially-trained models in a verifiable format such that the same can be stored over the blockchain and transferred from  $\mathcal{N}_1$ to $\mathcal{N}_2$. A naive idea for such representation is to condense the learning algorithm or the neural network model as a state machine or an automata (by representing each neuron or each layer of the model as a state or as a block in the blockchain~\cite{goel2019deepring}) which can then be governed by transition rules~\cite{hudson2019learning,schwartz2018sopa}, which can be translated to form a smart contract. However, such a formulation is not feasible since the number of neurons can be very large and it is not straightforward to represent the learning algorithm (backpropagation, gradient descent, etc.) in terms of state machine.

\subsection{Solution Overview}
\begin{figure}[!t]
    \centering
    \includegraphics[width=0.8\linewidth]{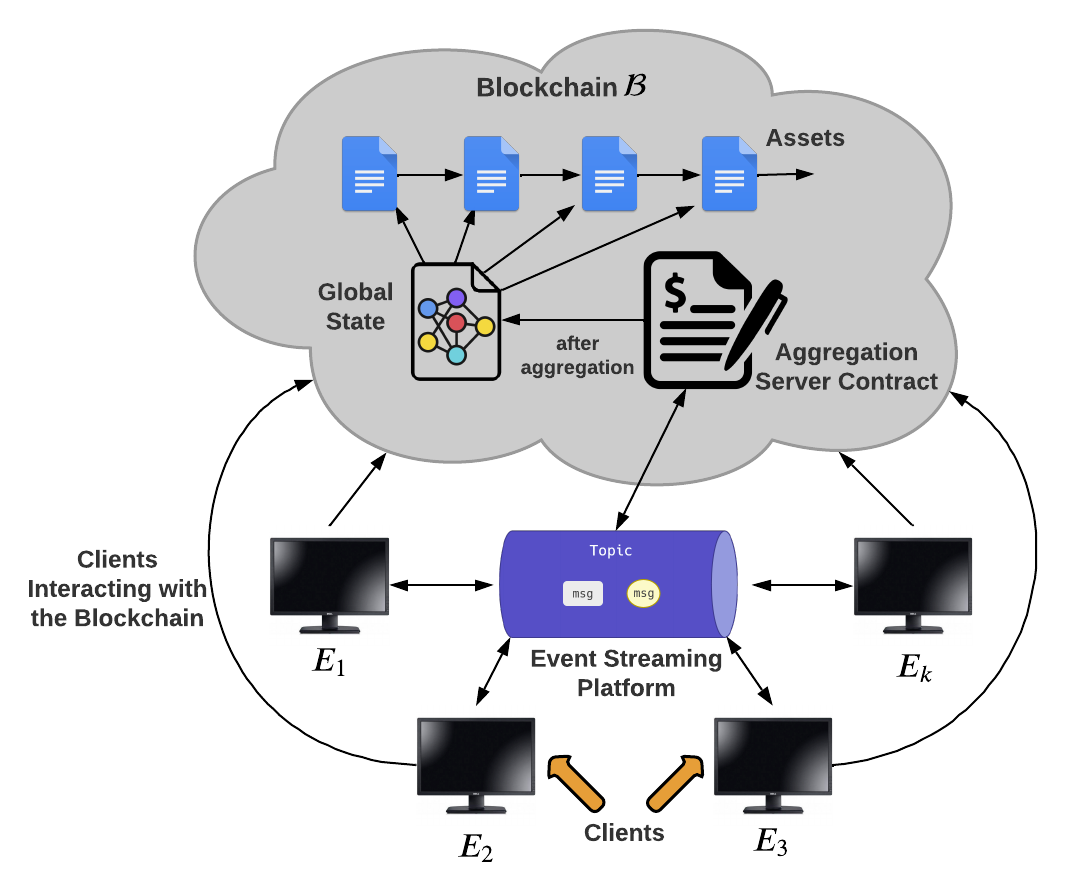}
    \caption{Figure shows a simplified design of the Federated Learning architecture and illustrates the working procedure of uploading the learning models in the form of asset from a federated network to the blockchain network.}
    \label{fig:blockchain}
\end{figure}
\ourmethod{} contains two primary components in its end-to-end architecture -- (1) An FL platform over individual networks $\mathcal{N}_1$ and $\mathcal{N}_2$ (as shown in Fig. \ref{fig:PoTF}) to train local models independently and update global model versions within each blockchain network, and (2) A relay service for verifiable transfer of learning assets from $\mathcal{N}_1$ to $\mathcal{N}_2$, and vice-versa.

Fig.~\ref{fig:blockchain} shows an overall view of the different architectural components of \ourmethod{}. At its core, we have a set of clients that own the data and maintain the local models. The clients are connected to a blockchain platform and fetches the latest version of the global model from the blockchain and update their local models by retraining them over the new data. These local models are forwarded to the blockchain, and the \textit{Aggregator Server Contract}, a smart contract to aggregate local models is triggered to aggregate them. \ourmethod{} uses Federated Averaging~\cite{mcmahan2017communication} for model aggregation, although any aggregation method can be used. Once the aggregated global model is updated in the blockchain, the next iteration of the FL starts. 

A critical aspect of this design is that the clients and the \textit{Aggregation Server Contract} need an ordering service over two-way communication, as the smart contract needs to update the global model after the clients forward their local models. Similarly, the clients should start the next iteration once the smart contract aggregates the local models and generate the global model. To synchronize the operations among the clients and the \textit{Aggregation Server Contract}, we use an ordering service using an event streaming platform.

\subsection{Event Streaming for Ordering Local/Global Models}
We have employed an \textit{event streaming} platform among the clients and the \textit{Aggregation Server Contract} in a publisher-subscriber setting for ordering different versions of the model. The clients use this streaming channel to publish the local models after each update. Once the local models from all the clients (or a predefined number of clients, to address stragglers) are available on the stream queue, the \textit{Aggregation Server Contract} is triggered to generate the global model by aggregating the local models. Once the aggregated model is available, it is published over the streaming channel, and the clients can subscribe to the message from the stream queue to look for the updated global models. For straggling clients that are not  up-to-date with the current updated model, the streaming platform provides flexibility to store multiple versions of the global model in the subscribed stream queue, from which the client can choose the latest/preferred version of the model. Though the streaming infrastructure provides a comprehensive platform to share global and local models, a couple of open challenges need to be addressed for executing the framework in practice -- (1) representation of learning assets within a blockchain and (2) cross-chain transfer of assets corresponding to global model versions for extending the federated training across multiple networks.

\section{Representation of Learning Assets} \label{sec:blockchain-FL}
As mentioned earlier, existing works~\cite{goel2019deepring,awan2019poster,ramanan2020baffle,peng2021vfchain,feng2021blockchain} primarily advocate for a structural embedding of a DNN as an asset to be stored in the blockchain. This section first analyzes why such structural embedding does not work when the learning assets need to be transferred from one blockchain to another. 

\subsection{Pilot Study -- Effect of Structural Embedding} \label{sec:pilot}
We first present an elementary pilot study to demonstrate that a structural embedding of the DNN as a learning asset is not suitable for cross-chain model training. For this purpose, we employed two learning CNN models with a similar backbone VGG structure. The two models differed in the convolution layers applied; while Model-1 used $3 \times 3$ convolution kernels, Model-2 is a compressed version of Model-1 and used Fire modules~\cite{iandola2016squeezenet} in place of convolution layers to reduce the number of model parameters. The models were trained on \textit{Cifar-10} dataset for a total of $300$ iterations. The batch size employed for the experiment was $16$ in both cases. We have used Hyperledger Fabric as the blockchain network with only $2$ client nodes (the minimum size of any network).

\begin{table}[!ht]
\centering
      \caption{Performance Measures for Inserting a FL Model in a Blockchain}
      \label{tab:pilot}
      \scriptsize
\begin{tabular}{|l|c|c|c|}
\hline
\multirow{2}{*}{} & \textbf{}       & \textbf{Model-1} & \textbf{Model-2} \\ \cline{2-4}
& Accuracy        & 53.92\%                     & 47.46\%                       \\ \hline
& \#activations   & 89866                       & 144394                        \\
\multirow{3}{*}{\shortstack[l]{\textbf{Structure Embedding}\\(~\cite{goel2019deepring,feng2021blockchain})}} & \#layers        & 15                          & 37                            \\
  & Asset Size & 98.24MB                     & 50.01MB                       \\
 & Insert Time     & 229.291 s                   & 102.452 s                     \\ \hline 
\multirow{3}{*}{\shortstack[l]{\textbf{Parameter Embedding}\\(\ourmethod)}} & \#parameters    & 865482                      & 188002                        \\
& Asset Size      & 19.9MB                      & 4.5MB                         \\
 & Insert Time     & 48.823 s                    & 9.693 s                       \\ \hline
\end{tabular}
\end{table}

The implications of the two models during model training are tabulated in Table \ref{tab:pilot}. In the table, the `\textit{Asset Size}' reported refers to the size of the JSON file storing the model representations, while `\textit{Insert Time}' is the time it takes to insert the asset structure into the blockchain. We consider two embeddings of the model for the representation of an asset in the blockchain. (a) \textbf{Structural embedding}: the pixel activation and the layers of the global model are embedded using existing approaches such as~\cite{goel2019deepring,feng2021blockchain}. (b) \textbf{Parameter Embedding}: the weights/gradients of the model learned during the training are embedded as the asset structure. From the table, we observe that structural embedding costs significantly in terms of asset size and the time required to insert the asset in the blockchain. With increasing size of the model, the number of activations and layers increases, and often becomes intractable. However, if parameters are used, which can be shared over few activations, or even layers, the asset size decreases. This change in asset size is evident in deeper models than in the shallow ones. This cost increases drastically when multiple versions of the global model need to be exchanged between blockchains for verifiablity purpose. Consequently, parameter embedding gives a better alternate for an asset representation and a compression technique like~\cite{iandola2016squeezenet} applied on the Model-2 helps reduce the asset size and insert time further.

\textbf{Problem with Parameter Embedding:} Although parameter embedding reduces the asset size significantly, one major limitation of it is that the parameters learned during an iteration displays partial information and  cannot alone ensure verifiability. Therefore, additional information must be included with the asset structure to verify that the model weights/gradients indeed provide a correct representation of the model. 

\subsection{Asset Representation with Model Parameters}
Based on the above analysis, the asset structure in \ourmethod{} represents additional parameters which are essential to verify the correctness of the model. The core idea is to use the standard logic of \textit{model validation}, as widely used during the development of DL models, for the verification purpose. Every network typically exposes a public validation dataset contributed by its clients, to validate the global model generated at each iteration. It can be noted that the idea of using an anonymized public validation dataset for FL model validation is well adopted in the existing literature~\cite{sheller2020federated,choudhury2020anonymizing} and also used for detecting attacks like model poisoning~\cite{bhagoji2019analyzing,fang2020local}. We adopt this idea for model parameter verification. Thus, the asset includes (a) the model parameters, i.e., the weights/gradient learned during an iteration,  (b) \textit{the model hyperparameters}, (c) \textit{random seed} and the \textit{optimizer} used, (d) a pointer to the \textit{public test dataset}, and (e) \textit{metrics} as observed by the client/updater during local/global model training and testing. Further, a learning asset is digitally signed by the client (for local model updates) or a set of clients as endorsers (for global model updates, details in Section~\ref{sec:cross-chain}). 

\subsection{Verification Logic} 
\ourmethod{} uses the public validation data to validate the model for verifying the learning assets. If the computed accuracy by executing the model with the stored hyperparameter configurations within the asset is less than the accuracy reported within the asset, the corresponding asset is discarded. We use the idea of model reproducibility~\cite{pineau2020building} here, which ensures that the model should not deviate from the reported accuracy and loss when executed with the same dataset with the same hyperparameter configurations. Apart from the logic verification, \ourmethod{} also verifies the digital signatures that endorse a particular asset representing a learning model.   

    
        
        

\begin{figure}[t]
    \centering
    \includegraphics[width=\linewidth]{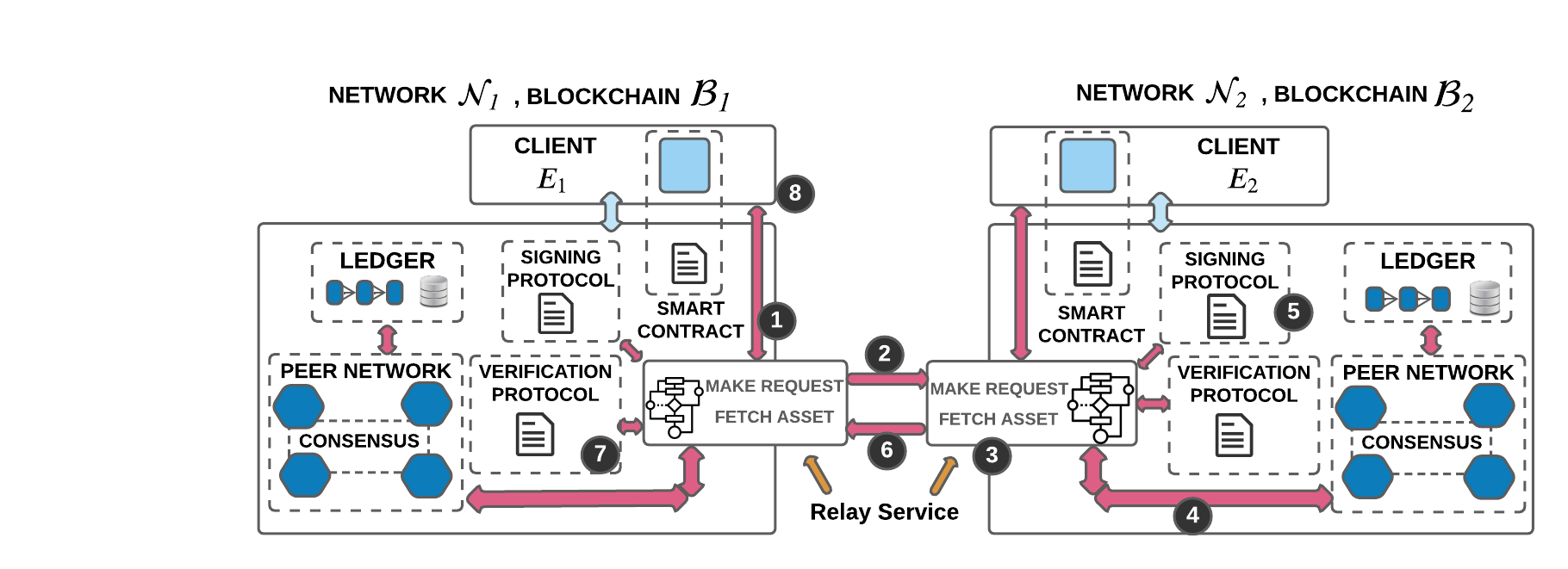}
    \caption{Control Flow for Cross-Chain Interoperation of Assets (denoted by red numbered arrows) between two permissioned Blockchain networks.}
    \label{fig:interoperation}
\end{figure}

\section{Cross-Chain Transfer of Learning Assets} \label{sec:cross-chain}
The final component of \ourmethod{} is a protocol for cross-chain asset transfer and its verification. Fig.~\ref{fig:interoperation} shows a schematic diagram explaining this process. Indeed, the design of \ourmethod{} makes this component simple, where we augment an existing interoperability architecture~\cite{abebe2019enabling} for asset exchange and verification in sync with the model updates. 

\subsection{Transfer Protocol - Asset Exchange}
Considering two networks $\mathcal{N}_1$ \& $\mathcal{N}_2$ and the corresponding blockchains $\mathcal{B}_1$ \& $\mathcal{B}_2$, the network clients initiate this asset exchange and work as a relay (by running a relay service) between the two networks. Considering the use-case as depicted in Fig.~\ref{fig:usecase}, one of the hospitals in the Hospitals' network and one of the diagnostic centers in the Diagnostic Centers' network can establish this relay communication to exchange the most updated learning assets between the two networks. We assume the existence of an \textit{Identity Interoperable Network} (IIN) as a blockchain service similar to~\cite{ghosh2021decentralized}, which helps the clients to access the identity (public key) of the other clients in a different network. Based on the identity information, client $E_1 \in \mathcal{N}_1$ requests for the most recent global model (or a specific version of the global model) from the client $E_2 \in \mathcal{N}_2$ through the local relay service of $\mathcal{N}_1$ (\textbf{Step 1}). The relay service of $\mathcal{N}_1$ then serializes the message received from $E_1$ and send it to the relay service of $\mathcal{N}_2$ (\textbf{Step 2}), where it is decoded to extract the parameters, like the model version, etc. (\textbf{Step 3}). The relay of $\mathcal{N}_2$ then initiates a transaction to $\mathcal{B}_2$ to access the most recent global model (or the requested version); the transaction goes through the local blockchain consensus of $\mathcal{B}_2$ (\textbf{Step 4}). To prevent the relay from exhibiting any Byzantine behavior, we trigger a blockchain transaction and pass it through a local consensus rather than directly fetching the model from the blockchain.

In response to the transaction against a cross-chain access request, the blockchain $\mathcal{B}_2$ triggers a local service (a smart contract). Through this service, the requested asset (the global model corresponding to the version requested) passes through a Byzantine agreement based on the \textit{Collective Signing} (CoSi)~\cite{syta2016keeping,kogias2016enhancing} protocol (\textbf{Step 5}). In CoSi, the majority of the peers collectively sign the asset using the Boneh-Lynn-Shacham (BLS)~\cite{boneh2001short} cryptosystem to ensure that the correct asset is being transferred to the other network; the details of the protocol can be found in~\cite{chander2019fault}. Finally, the relay of $\mathcal{N}_2$ communicates the signed asset to the corresponding relay of $\mathcal{N}_1$ (\textbf{Step 6}) along with the verification credentials (public keys of the signees). Finally, after Byzantine agreement to verify the asset received from $\mathcal{N}_2$ (\textbf{Step 7}), the client application $E_1 \in \mathcal{N}_1$ updates its learning asset to its local blockchain (\textbf{Step 8}). We next discuss the verification process.

\subsection{Verification Process}
The verification protocol needs to verify two things -- (1) the received learning asset has passed through the CoSi-based Byzantine agreement from $\mathcal{N}_2$, and (2) the asset contains the correct model parameters. For (1), the clients use the IIN to fetch the public keys of $\mathcal{N}_2$'s clients and use the CoSi verification~\cite{syta2016keeping} using the BLS cryptosystem. As noted in~\cite{chander2019fault}, CoSi verification using BLS signatures is extremely fast and scalable; therefore, this verification round is much light-weight. For (2), the clients uses the \textit{Verification Logic} as discussed in Section ~\ref{sec:blockchain-FL} to verify the received model weights using the additional parameters from the received learning asset. One critical aspect is the access to the public validation data maintained by the clients of $\mathcal{N}_2$. There can be multiple solutions, like (i) use the same Byzantine agreement protocol to transfer a hash and a pointer of the public validation data from $\mathcal{N}_2$ to $\mathcal{N}_1$, or (ii) use the IIN to access a pointer to the public validation data. In our implementation, we use the second approach. 

\subsection{Cross-chain Training} 
As \ourmethod{} uses an ordering service to store and retrieve the models from the blockchain, this step is pretty straightforward. Once a pre-trained global model from $\mathcal{N}_2$ is available on $\mathcal{B}_1$, the clients of $\mathcal{N}_1$ can use that global model to update their local weights and trigger \textit{Aggregation Server Contract} for the global model update. The \textit{Aggregation Server Contract} can then aggregate the local models to construct a new version of the global model capturing the learned parameters from the clients of both networks.

%% file: tex/05.Implementation.tex
\section{Experimental Setup} \label{sec:implementation}
We have implemented \ourmethod{} as a standalone toolbox and tested it thoroughly over large networked systems deployed through multiple container networks, with $10$ to $40$ Docker containers representing FL clients. The entire experiments have been executed over $9$ Amazon EC2 T2.2Xlarge instances having octa-core CPU with 32GB memory running on 3GHz Xeon processors. Each of these EC2 instances hosts multiple Docker containers restricted to a single CPU-core and a maximum of $2$GB memory, with the associated federated training service running over them. 


\subsection{Design Specifics}
We use the Hyperledger Fabric\footnote{\url{https://www.hyperledger.org/use/fabric} (Access: \today)} version 2.2.0 to implement the blockchain networks. Every docker container runs one Fabric client service to connect to the blockchain network. The containers execute the FL client module and use the Fabric API to initiate transactions for storing or retrieving the assets. We use an overlay network based Docker Swarm\footnote{\url{https://docs.docker.com/engine/swarm/} (Access: \today)} spanned over the EC2 instances to interconnect the containers. We kept the network bandwidth between two containers in a Docker overlay network within $1$ to $5$ Gbps, which is the typical minimum bandwidth used in enterprise networks. Each silo has a separate swarm, where the containers within each swarm use a single Fabric overlay network. 

The \textit{Fabric-go-sdk} limits the size of the byte array encoded within a single transaction to around $1.2$MB. However, the size of the \ourmethod{} learning assets vary depending on the number of parameters used in the model. Therefore, we segment a learning asset into multiple fragments of $800$KB each. Since each asset must be stored with a unique ID, we store each fragment with an ID $\lbrace$Asset ID, Fragment Number$\rbrace$. Consequently, during the cross-chain asset transfer, the relay service retrieves all the fragments of an asset and then stitches them to form a single asset. We used Kafka\footnote{\url{https://hyperledger-fabric.readthedocs.io/en/release-2.2/kafka.html} (Access: \today)} publish/subscribe platform for event streaming that runs the ordering service over the Fabric clients.

\subsection{Dataset and Learning Models}
To train the learning models via FL, we have used \textit{Cifar-10} dataset~\cite{krizhevsky2009learning} containing 50k training and 10k test images; image sizes are $32 \times 32$ and are divided into ten classes. The complete training dataset is distributed identically and independently (i.i.d) among the FL clients, such that each client has an almost equal number of images from each class. The clients then use their respective datasets for training with a train-test split of $0.1$. Hence, the entire dataset has three logical partitions: a local train set \& a local test set for each client and a global test set. To evaluate the effectiveness of the cross-chain transfer training, we perform an image classification task on various models using \textit{Cifar-100} dataset~\cite{krizhevsky2009learning}, which is similar to Cifar-10, but the images are divided among $100$ fine categories and $20$ coarse sub-categories. We use the coarse categories for labeling and evaluation.

We use four different models to evaluate the performance of \ourmethod{} -- (1) \textbf{SimpCNN}, a $6$-layer convolution neural network (CNN) model ($3 \times 3$ kernel) with the structure of \textit{conv32-conv32-pool-conv64-conv64-pool-conv128-conv128-pool} followed by a feed-forward dense hidden unit of 256 neurons and an output layer, (2) \textbf{CompVGG}, a compressed version of the VGG-11~\cite{simonyan2014very} model having a total of $7$ \textit{``convolution"} layers with a backbone of \textit{conv32-pool-conv64-pool-conv128-conv128-pool-conv128-conv128-conv128-pool} where the \textit{convolution} layers are replaced by the \textit{Fire module} inspired by SqueezeNet~\cite{iandola2016squeezenet}, (3) \textbf{MobileNet-V2}~\cite{howard2017mobilenets}, a practical large scale CNN for mobile visual applications, and (4) \textbf{Resnet-18}~\cite{hara2018can}, an 18-layered large-scale CNN model used in many practical visual applications. MobileNet-V2 and Restnet-18 are particularly used to analyze the cross-chain model transfer overhead for large models. 

\subsection{Hyperparameters Tuning and Performance Metrics}
We use synchronous FL training, where the version of the global model is updated after every global round. Each global round includes $2$ local iterations. For the model training, we have used Sparse Categorical Cross-Entropy as the loss and Adam Optimizer with a learning rate of $0.001$.

To record intra-chain performances, we measure the \textit{Accuracy} and \textit{Loss} of the learning models for evaluating their performance at four different points of execution. (i) \textbf{Pre-Test} is the metric recorded on the global model on the global testset. (ii) \textbf{Pre-Val} is recorded on the global model on each client testset, averaged over all the clients. (iii) \textbf{Post-Test } is the metric value recorded on the individual client models on the global testset, averaged over the number of clients. (iv) \textbf{Post-Val} is recorded on the individual client models on each client testset, averaged over the number of clients. To analyze the overhead during cross-chain asset transfer, we report the latency incurred to create an asset, the asset retrieval time, the time needed for CoSi-based Byzantine agreement, and finally, the time for model verification. 

\textbf{Unavailability of Baselines}: It can be noted that to the best of our knowledge, \ourmethod{} is the first of its kind that proposes a cross-chain model training framework. As we have shown in Table~\ref{tab:pilot} under Section~\ref{sec:blockchain-FL}, existing works for accountable FL are not suitable for cross-chain training, as they use a structural embedding of the model to represent an asset. So, we do not compare the performance of \ourmethod{} with other existing approaches to avoid unfair comparison.

%% file: tex/06.Evaluation.tex
\section{Evaluation}
We evaluate \ourmethod{} from three different aspects -- (a) the overall performance of the FL system, (b) the efficacy of the transferred weights, and (c) the transfer overheads. 

\subsection{Performance of Federated Learning  System} \label{sec:performance_fl}

\begin{figure}[ht!]
 \centering
   \subfigure[\footnotesize{\textit{SimpCNN - 10 Clients}}]{\includegraphics[width=0.47\linewidth]{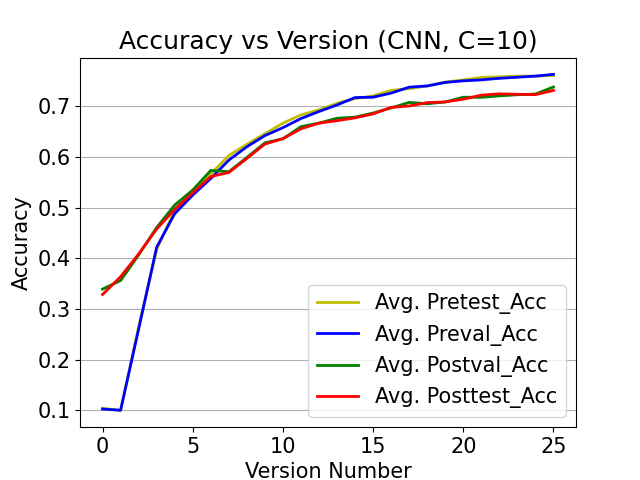}}
   \subfigure[\footnotesize{\textit{CompVGG - 10 Clients}}]{\includegraphics[width =0.47\linewidth]{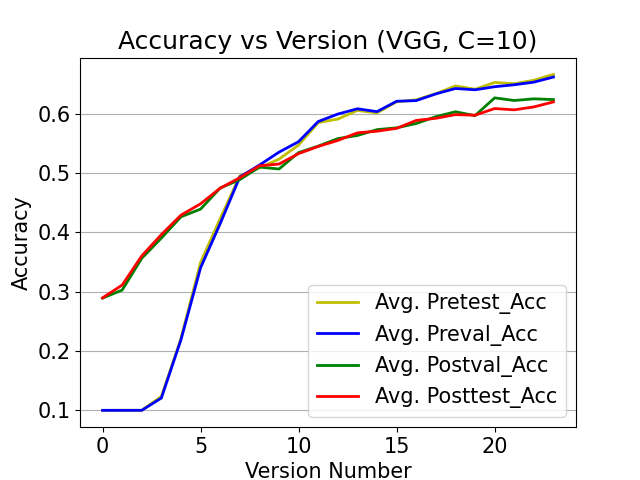}}
   \subfigure[\footnotesize{\textit{SimpCNN - 40 Clients}}]{\includegraphics[width=0.47\linewidth]{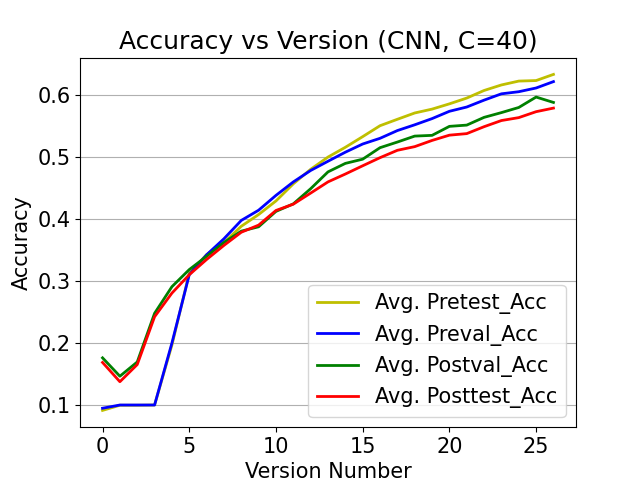}}
   \subfigure[\footnotesize{\textit{CompVGG - 40 Clients}}]{\includegraphics[width =0.47\linewidth]{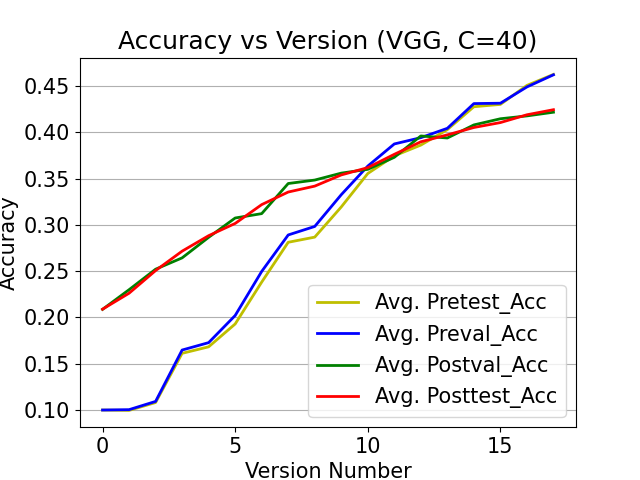}}
\caption{Comparison of accuracy metrics for both models against version numbers}
\label{fig:all_acc}
\end{figure}

With our main motivation being cross-chain transfer and the validity of the weights transferred, we evaluate how FL performance changes on varying settings, and not the validity of FL itself by comparing with centralized training. Fig.~\ref{fig:all_acc} shows the accuracy metrics as explained earlier for the two models -- SimpCNN and CompVGG. We employed a batch size of 32 for both models. From the figure, we observe that the Pre-Test accuracy is higher than the Post-Test accuracy; that is, the accuracy of the aggregated global models surpasses that of the individual trained local models. This shows that aggregation of the weights via federated averaging results in better learning of the model. The result is consistent across different number of clients. To explain this behavior, we hypothesize that aggregation provides a regularizing effect since no client model gets overfitted on their local dataset, hence, the model accuracies improve on aggregation than just on the individual learned model. We observe that SimpCNN exhibits higher accuracy than CompVGG since the former is a comparatively larger model (details in Table \ref{tab:asset_entry}). It is to be noted that the version number is incremented after the \textit{Aggregation Server Contract} aggregates the local models received from the clients. 

However, we observe that when trained with 10 clients, the model exhibits higher accuracy as compared to with 40 clients. Interestingly, we notice that the average global round duration for both the models is lower for the latter (3 min 23 sec in CompVGG and 4 min 39 sec for SimpCNN). This is because with a total of 50k training images, and an increase in the number of clients, each client receives a lower number of local data points. Hence, the local model takes less duration for an epoch but overfits the dataset, lowering the efficacy.

\subsection{Evaluating Efficacy of Transferred Weights}
\begin{figure}[ht!]
 \centering
   \subfigure[\footnotesize{ }]{\includegraphics[width=0.49\linewidth]{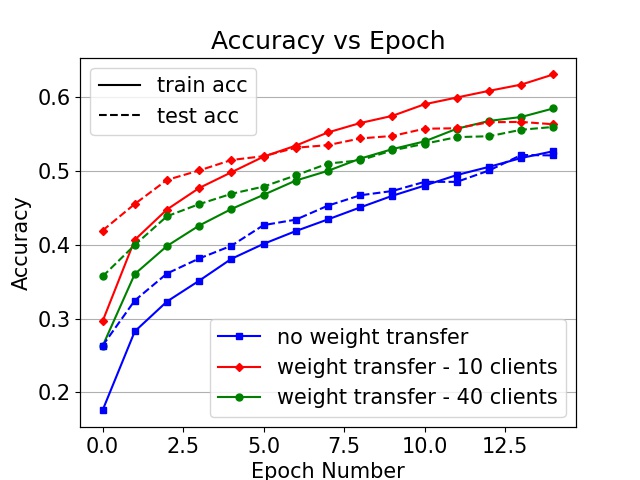}}
   \subfigure[\footnotesize{ }]{\includegraphics[width =0.49\linewidth]{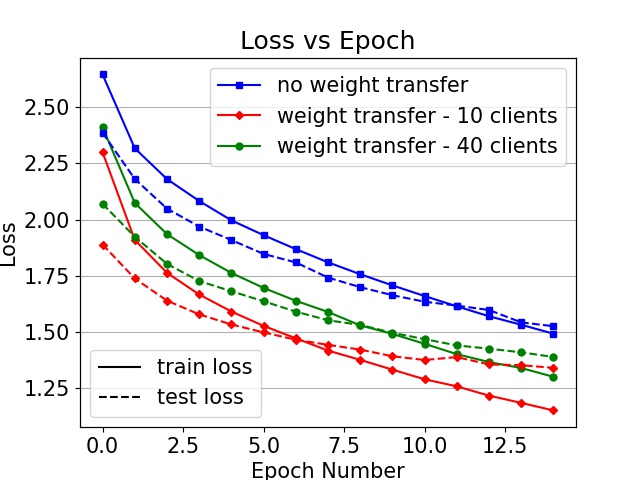}}
\caption{Accuracy and Loss(Training and Testing) of SimpCNN on Cifar-100 dataset trained from scratch as well as initializing the model with the transferred weights.}
\label{fig:transfer_same}
\end{figure}

After the successful transfer of signed learning assets among blockchain networks, the receiving network can use the weights to train a learning model on possibly, a different dataset. Here, we experiment the efficacy of the transferred weights through a transfer learning task. As already established, we retrain an image classification model on the Cifar-100 dataset with coarse labels as the targets. We first report how the transferred weights perform as an initialization of the SimpCNN model, that helps us understand how well the weights already trained for the same model work on a different dataset. As a baseline, we train SimpCNN from scratch, with random initialization, and no weight transfer. We plot the results for the same in Fig.~\ref{fig:transfer_same}.

We notice that the results of training and testing accuracy of SimpCNN on the Cifar-100 dataset resonates with the performance of SimpCNN on the Cifar-10 dataset. Similar to the observation reported in Section \ref{sec:performance_fl}, SimpCNN performs better when it was initialized with the transferred weights trained on the model via a federated setup of $10$ clients than when the initialization was done with the weights trained via federated setup for $40$ clients. This behaviour is precisely what we earlier encountered in Subsection~\ref{sec:performance_fl}. Hence, the quality of weights transferred remains superior as well, establishing the correctness of the weights transferred. Nonetheless, the model trained with the shared weights performed better than the model trained from scratch, confirming the efficacy of the learning model with cross-chain training.

\textbf{Impact on Model Augmentation}: Additionally, we run the same transfer learning task but on a different model having the first few layers (body) same as SimpCNN, but the \textit{head} with three additional $3 \times 3$ convolution layers of 256 filters, followed by a max-pooling layer, a dense layer of 512 units and finally the output layer. This experiment essentially establishes the insights using the transferred weights as initialization to a different model where only a few layers can be initialized. We call this model \textit{TransferCNN}. 

\begin{figure}[ht!]
 \centering
   \subfigure[\footnotesize{ }]{\includegraphics[width=0.49\linewidth]{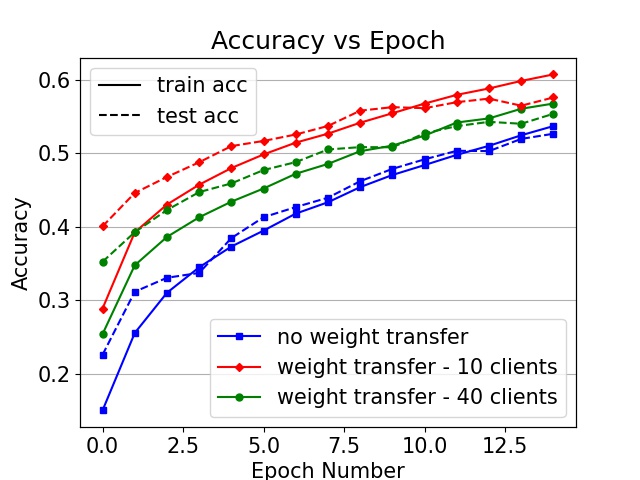}}
   \subfigure[\footnotesize{ }]{\includegraphics[width =0.49\linewidth]{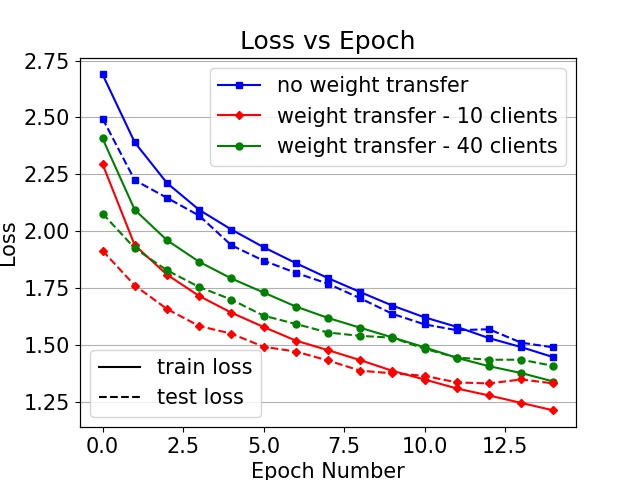}}
\caption{Accuracy and Loss(Training and Testing) of TransferCNN on Cifar-100 dataset trained from scratch as well as initializing the model with the transferred weights.}
\label{fig:transfer_diff}
\end{figure}

On a similar note, we observe in Fig.~\ref{fig:transfer_diff} that TransferCNN performs better when initialized with the weights learned during a federated setup with $10$ clients than when the initialization was done with the weights trained via federated setup for $40$ clients. However, in both cases, the accuracy achieved is higher than when TransferCNN was trained from scratch, without any transferred weight initialization. Thus, the experiments mentioned above establish the validity of the cross-chain transfer module and shows that the transfer of weights produces influential initialization variables for transfer learning.

\subsection{Analysis of Transfer Overheads}
To transfer the requested assets between blockchain networks, it incurs the following costs from the intermediate steps -- (a) cost for creating a learning asset from the model parameters and include it to the Fabric in multiple fragments, (ii) cost for retrieving an asset from the ledger on request and defragmenting it, (iii) cost for the Byzantine agreement protocol to collectively sign the asset, and (iv) verification of the asset after the transfer is complete. 

\subsubsection{Entering Asset}
Table \ref{tab:asset_entry} shows the overhead required to embed the weights in the form of assets and to store them into the ledger after fragmenting. The values that we record are averaged across executions of $5$ runs. We observe that the asset size and the asset entry time scales up linearly  with an increase in the model dimension, as expected. However, it is to be noted that the time taken to enter the asset into the ledger by the FL server is around $1$ minute, even for a 232 MB sized asset (for ResNet model), while, the global round duration is atleast $3$ minutes even for SimpCNN. Therefore, \ourmethod{} prevents any contention, since an asset will be entered before the asset for the next round arrives, and the clients can avail the latest version of the global model well before an updated version of the same is generated. 

\begin{table}[h]
\centering
\caption{Overhead for Asset Creation}
\scriptsize
\begin{tabular}{|c|c|c|c|c|}
\hline
\textbf{Metrics} & \textbf{CompVGG} & \textbf{SimpCNN} & \textbf{MobileNet} & \textbf{ResNet} \\ \hline
\begin{tabular}[c]{@{}c@{}}\# params\\ \phantom{ }\end{tabular} & 171,682 & 814,122 & 3,239,114 & 11,192,019 \\ \hline
\begin{tabular}[c]{@{}c@{}}Asset Size\\ (MB)\end{tabular} & 4.0 & 19.5 & 67.3 & 232.1 \\ \hline
\begin{tabular}[c]{@{}c@{}}\# chunks\\ \phantom{ } \end{tabular} & 5 & 24 & 84 & 290 \\ \hline
\begin{tabular}[c]{@{}c@{}}Entry Time\\ (sec)\end{tabular} & 1.148 & 5.527 & 17.869 & 63.079 \\ \hline
\end{tabular}
\label{tab:asset_entry}
\end{table}

\subsubsection{Transferring Asset}
On an asset transfer request arrival at a relay service, it must first retrieve the requested asset from the ledger according to Steps 4--7 mentioned in Section \ref{sec:cross-chain} and defragment it. It then acts as an initiator to order the witness cosigners to sign the asset using BLS signatures and finally replies with the signed asset. The relay service must verify the signature on the received asset on the receiving end before committing the transferred asset to its local ledger. In Table \ref{tab:transfer}, we illustrate the timing overhead incurred at each of these steps. Retrieval Time includes the total amount of time in seconds to retrieve the asset as well as defragment it.

\begin{table}[h]
\centering
\caption{Overhead for Cross-chain Asset Transfer}
\scriptsize 
\begin{tabular}{|c|c|c|c|c|}
\hline
\textbf{Metrics} & \textbf{CompVGG} & \textbf{SimpCNN} & \textbf{MobileNet} & \textbf{ResNet} \\ \hline
\begin{tabular}[c]{@{}c@{}}Asset Size\\ (MB)\end{tabular} & 4.0 & 19.5 & 67.3 & 232.1 \\ \hline
\begin{tabular}[c]{@{}c@{}}Retrieval Time\\ (sec)\end{tabular} & 0.404 & 2.261 & 6.584 & 22.051 \\ \hline
\begin{tabular}[c]{@{}c@{}}CoSi Time\\ (sec)\end{tabular} & 0.847 & 5.241 & 18.586 & 110.460 \\ \hline
\begin{tabular}[c]{@{}c@{}}Verification Time\\ (sec)\end{tabular} & 0.871 & 6.017 & 19.844 & 154.705 \\ \hline
\end{tabular}
\label{tab:transfer}
\end{table}

As can be inferred from the table, the transfer time of the assets steadily increases with increasing model complexity. However, unlike the trend in Table~\ref{tab:asset_entry}, the cost of transfer overhead scales slightly faster than linear, as is evident from the increase in signing and verification times for ResNet. In the experiment above, we have transferred the asset via the HTTP POST request-response mechanism. To alleviate the increasing transfer requests while scaling up the model, techniques involving segregating the data channel from the control channel and spawning multiple processes to handle requests from relay services of different blockchain networks are promising fronts; however, these are not in the scope of the current work.

%% file: tex/07.Conclusion.tex
\section{Conclusion}
This paper presented an end-to-end framework that can learn a model and store it in a blockchain, which can then be transferred to other blockchain networks on-demand, thus increasing the scope and efficacy of the model training over multiple networks with rich and diverse training datasets. We constructed a robust federated learning system that can leverage various enterprises as FL clients and train the model on them. Additionally, the federated learning system stores an asset constructed from the model parameters after each global round in the blockchain network, which is transferable with the support of independent asset verification. Our extensive experimentation deciphers the efficacy of the transferred weights on the receiving end as well as the overhead costs incurred. 

A critical aspect of our framework is that it can leverage the global models trained over a different network and use the learned weights to initialize another model having a partially similar structure, an approach well known to the DL community based on transfer learning. As we observed during the evaluation, a model initialization approach like this significantly boosts up the efficacy of the final model. Although transfer learning is instrumental in generating rich and compelling models~\cite{fang2013transfer}, it is seldom adopted in practice as transferring models across networks involve the possibility of various attacks like model poisoning. In this context, \ourmethod{} can enable the development of rich DL models trained over diverse datasets across different networks, albeit without explicitly exposing the datasets to the public space and eliminating the possibility of model poisoning.